\documentclass[usenatbib,useAMS]{mn2e} 
\pdfoutput=1
\usepackage[usenames,dvipsnames]{color}
\usepackage{graphicx}
\usepackage{amsmath}
\usepackage{comment} 

\newcommand{\figdir}{figures}

\newcommand{\ri}{\rmn{i}}

\newcommand{\fract}[2]{\leavevmode\kern.1em\raise.5ex\hbox{\the\scriptfont0 #1}\kern-.1em/\kern-.15em\lower.25ex\hbox{\the\scriptfont0 #2}}

\newcommand{\boldb}{{{\bmath{b}}}}
\newcommand{\e}{{\hat{\bmath{e}}}}

\newcommand{\boldk}{{\bmath{k}}}

\newcommand{\E}{{{\bmath{E}}}}
\newcommand{\F}{{{\bmath{F}}}}

\newcommand{\boldv}{{\bmath{v}}}

\newcommand{\vdot}{{\bmath{\cdot}}}

\newcommand{\B}{{\bmath{B}}}
\newcommand{\vcross}{{\bmath{\times}}}

\begin{document}


\title{Seismology of the Wounded Sun}

\author[Paul S.~Cally and Hamed Moradi]{Paul S. Cally\thanks{E-mail: paul.cally@monash.edu}
and
Hamed Moradi\thanks{E-mail: hamed.moradi@monash.edu}\\
Monash Centre for Astrophysics and
School of Mathematical Sciences,
Monash University, Victoria, Australia 3800}

\maketitle


\begin{abstract}
Active regions are open wounds in the Sun's surface. Seismic oscillations from the interior pass through them into the atmosphere, changing their nature in the process to fast and slow magneto-acoustic waves. The fast waves then partially reflect and partially mode convert to upgoing and downgoing Alfv\'en waves. The reflected fast and downgoing Alfv\'en waves then re-enter the interior through the active regions that spawned them, infecting the surface seismology with signatures of the atmosphere. Using numerical simulations of waves in uniform magnetic fields, we calculate the upward acoustic and Alfv\'enic losses in the atmosphere as functions of field inclination and wave orientation as well as the Time-Distance `travel time' perturbations, and show that they are related. Travel time perturbations relative to quiet Sun can exceed 40 seconds in 1 kG magnetic field. It is concluded that active region seismology is indeed significantly infected by waves leaving and re-entering the interior through magnetic wounds, with differing travel times depending on the orientation of the wave vector relative to the magnetic field. {This presages a new directional-time-distance seismology.}
\end{abstract}


\section{Introduction}

Classical local helioseismology \citep{GizBir05aa} uses Doppler velocity or intensity data from near the solar surface to probe the Sun's interior: thermal, magnetic, and flow properties may all be inferred with varying degrees of reliability. The overlying atmosphere, the chromosphere and corona, are generally assumed to not be important in this process, though a separate `coronal seismology' \citep{NakVer05aa,De-05aa,SteZaiNak12aa} using observations from coronal heights and aimed at determining the physical characteristics of coronal structures has developed in recent years. It is well-appreciated that the Sun's internal waves manifest in the atmosphere as well, particularly in active regions \citep{BogJud06aa,JefMcIArm06aa,KhoCal13aa}, but any back-reaction on the interior seismology has heretofore been largely ignored \citep[though see][for an explanation of acoustic halos that relies on reflected fast waves]{KhoCol09aa}. We argue that this neglect can lead to serious errors in interpretation of seismic data.

\begin{figure}
\begin{center}
\includegraphics[width=\hsize]{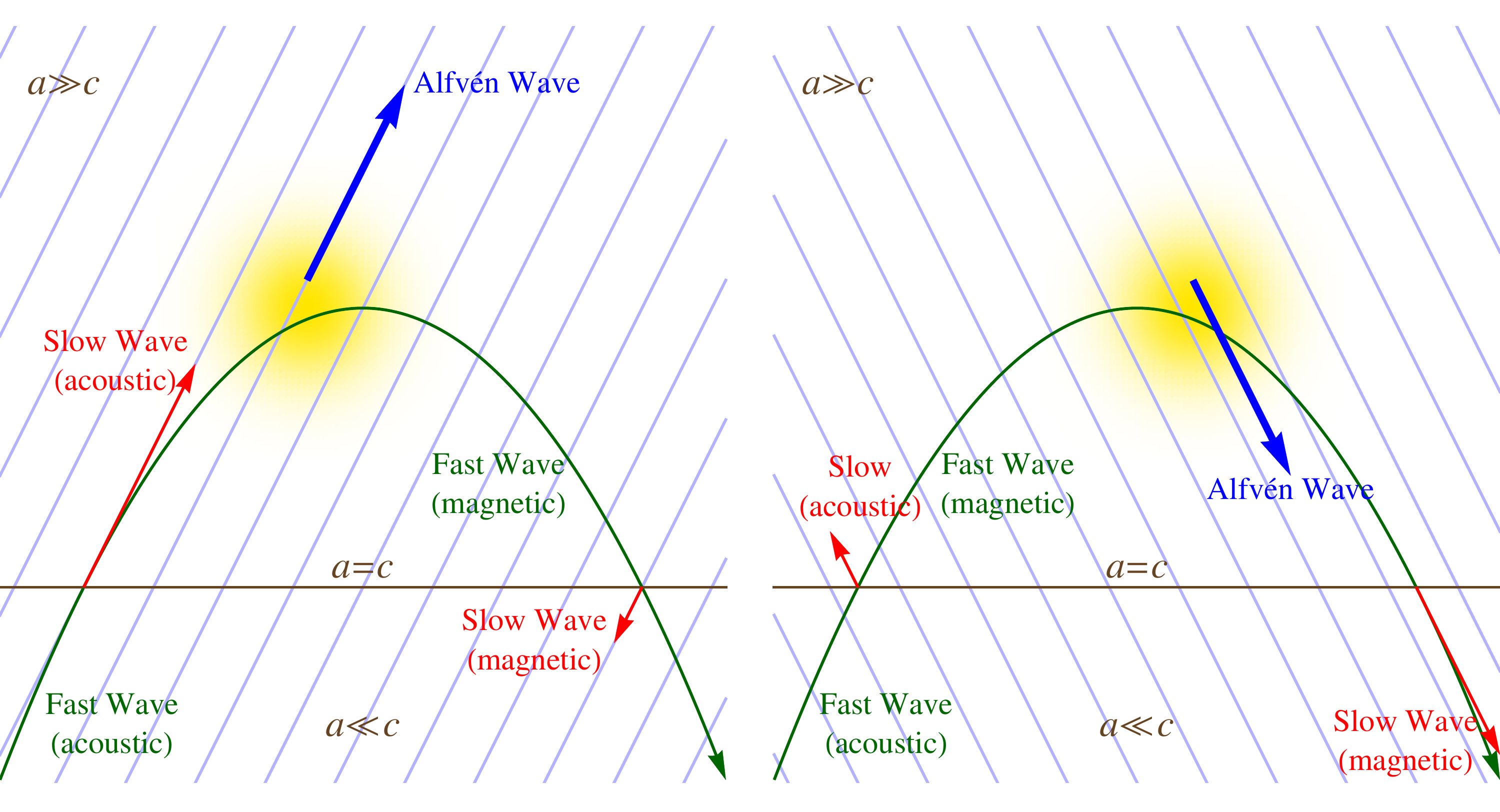}
\caption{Schematic diagram ($x$-$z$ plane) depicting fast/slow conversion/transmission at the equipartition level $a=c$, and fast-to-Alfv\'en conversion in a nebulous region near the fast wave reflection height. The magnetic field lines are shown as background. Conversion to upward or downward propagating Alfv\'en waves depends in the relative orientation of the magnetic field and the wave propagation direction. Both panels should be regarded as projections, with the fast wave raypath having a component in the $y$-direction; otherwise, there is no fast-Alfv\'en coupling. (Adapted from \citet{KhoCal12aa}) \label{fig:Alfschem}}
\end{center}
\end{figure}

The most widely used local helioseismic technique is Time-Distance Helioseismology \citep[TD;][]{DuvJefHar93aa,GizBir05aa}. By correlating observations of perturbations at different times and positions on the solar surface, a causative link is inferred and a travel time between pairs of points determined. Comparing these travel times with those calculated for a standard model, one infers the presence of wave-speed anomalies beneath the surface that may be due to such features as temperature variations or flows. Utilizing the full gamut of wave paths sampled by one's data set, inversions may be performed for subsurface structure. However, `travel time' is not necessarily as straightforward a concept as we might hope. Proper group travel time is difficult to measure with any precision, and in practice phase variations are used as a travel time proxy. 

This approach seems well-founded in quiet Sun, where magnetic effects may be ignored, but is subject to several uncertainties and complications in and around active regions dominated by magnetism in the surface layers and above. To set the scene, the quiet Sun's internal $p$-modes ($p$ for pressure) are essentially acoustic waves travelling in an unmagnetised stratified medium, with dispersion relation $\omega^2-\omega_c^2-c^2k_h^2=c^2k_z^2$, where $\omega$ is the frequency, $c$ is the sound speed (which increases with depth), $k_h$ and $k_z$ are the horizontal and vertical components of the wave vector, and $\omega_c$ is the acoustic cutoff frequency \citep[commonly defined by $\omega_c^2=(c^2/4H^2)(1-2\, dH/dz)$ where $H$ is the density scale height, but see][for a critique of such formul\ae]{SchFle98aa}. Vertical propagation is therefore limited to the cavity where the left hand side is positive. At depth, where $c^2k_h^2\gg\omega_c^2$ typically, the lower turning point is around the Lamb depth $\omega^2=c^2 k_h^2$, whilst at the upper turning point $\omega^2=\omega_c^2+c^2 k_h^2\geq\omega_c^2$. The introduction of strong magnetic field in surface layers has several effects to alter this picture:
\begin{enumerate}
\item Magnetic field alters the density and thermal structure of the plasma by supplying additional magnetic forces, requiring the plasma pressure to adjust accordingly (the `indirect' effect). Changes in the sound speed naturally affect wave travel times.
\item Magnetohydrodynamic (MHD) wave modes become available: fast, slow, and Alfv\'en. At depth, where the Alfv\'en speed $a$ is much less than the sound speed $c$, the $p$-modes are effectively fast waves. But on passing through the equipartition layer $a=c$ they partially transmit as acoustic waves (now slow) and partially convert to magnetically dominated fast waves \citep{Cal06aa,SchCal06aa,Cal07aa}, depending on the attack angle between the wave vector and the magnetic field. In sunspot umbrae, $a=c$ is typically situated several hundred kilometres below the surface, whilst in penumbrae it is around the surface \citep{MatSolLag04aa}. In the regions surrounding sunspots, $a=c$ may crudely be equated with the level of the magnetic canopy in the low atmosphere.
\item The transmitted slow waves above $a=c$ are essentially field-guided acoustic waves. Unlike their non-magnetic cousins though, they may still propagate vertically at frequencies below the acoustic cutoff (around 5 mHz), provided $\omega>\omega_c\cos\theta$, where $\theta$ is the field inclination from the vertical \citep{BelLer77aa}; the so-called `magneto-acoustic portals' \citep{JefMcIArm06aa}. This ramp effect has recently been shown computationally to be far more important than radiation in modifying the acoustic cutoff \citep{HegHanDe-11aa}.
\item Whether or not the slow wave escapes into the solar atmosphere in strong field regions, the fast wave certainly does. It is immune to the acoustic cutoff effect, but nevertheless has its own upper turning point substantially higher in the atmosphere. For simplicity, assume that this is high enough that $a\gg c$ and therefore ignore sound speed altogether. The dispersion relation is then $\omega^2=a^2(k_h^2+k_z^2)$, indicating reflection where $\omega^2=a^2k_h^2$. This is an important point: the fast wave reflects where its horizontal phase speed $\omega/k_h$ coincides with the local Alfv\'en speed. After it reflects, the fast wave re-enters the solar interior wave field, and therefore is part of the seismology of the Sun. Its journey through the atmosphere must therefore have some effect on wave timings.
\end{enumerate}

However, that is not the full story. The fast wave itself may partially mode-convert to upward or downward propagating Alfv\'en waves, depending on magnetic field orientation relative to the wave vector \citep{CalGoo08aa,CalHan11aa,HanCal12aa,KhoCal11aa,KhoCal12aa,Fel12aa}, thereby removing energy from the seismic field and potentially altering its phase (see the schematic diagram Figure \ref{fig:Alfschem}). This conversion typically occurs in a broad region near the upper turning point of the fast wave, but only if the wave is propagating across the vertical plane containing the magnetic field lines; we let $\phi$ denote the angle between the vertical magnetic flux and wave propagation planes. Such a phase change could be wrongly interpreted in TD as a travel time shift \citep{Cal09ab,Cal09aa}.

So, there are many processes going on here: fast/slow conversion/transmission of upward travelling waves at $a=c$; slow wave reflection or transmission, depending on the acoustic cutoff and the magnetic field inclination $\theta$; fast-to-Alfv\'en conversion near the fast wave turning point; partial reflection of the fast wave to re-enter the seismic field; further fast/slow conversion/transmission as the fast wave passes downward through $a=c$. In totality, this is too complicated to model analytically. However, knowing the physics of the constituent parts, a full numerical simulation can provide valuable insight. The novel aspect explored here is to simultaneously calculate the TD travel time shifts $\delta\tau$ and upward acoustic and Alfv\'enic losses in a simulation as functions of $\theta$ and $\phi$ (for selected frequencies and horizontal wavenumbers).

The considerations set out above indicate that active regions are truly wounds in the surface of the Sun. They allow waves and wave energy to escape from the interior cavity wherein they are normally trapped, but also allow their depleted, phase-shifted, and mode-converted remnants to re-enter and infect the local seismology that we rely on for information about the sub-surface. The purpose of this paper is to look for correspondences between acoustic and magnetic wave losses on the one hand, and `travel time' shifts on the other, and thereby to verify that the seismology of the wounded Sun depends on propagation, transmission, and conversion in overlying active region atmospheres. {We introduce a directionally filtered time-distance approach\footnote{A  wedge filter was suggested for use in TD by \citet{Gil00aa} to measure either rotation or meridional circulation; see his fig.~4.4. \citet{ChoLiaYan09aa} employed a similar idea in the construction of an acoustic power map.} to simulation data that is directly extensible to real helioseismic data, and thereby envisage a Directional-Time-Distance (DTD) seismology sensitive to magnetic field orientation. In Section \ref{sec:BVP}, the DTD results are verified using the Boundary Value approach of \citet{Cal09ab}.}

{ The purpose of this paper is to demonstrate that the above mode conversion scenarios actually operate in the solar context; to survey how they depend on field inclination $\theta$ and wave orientation $\phi$; and to assess the utility of DTD in discerning directional magnetic effects. To this end, we explore only vertical atmospheric stratification, characterized by density scale heights of order 0.15 Mm, and ignore magnetic field inhomogeneity, which in sunspots has scale length of order many Mm. A subsequent paper will address wave simulations in realistic sunspot models using DTD. This is not to say though that flux tubes are not important to wave propagation, though probably they are more relevant to the corona where the far greater gravitational scale height is no longer the primary feature, leaving complex loop structures to dominate \citep{VanBraVer08aa}. Chromospheric sunspot fields, dominated be a single monopolar spot, may be expected to be smoother.
}

\section{Computational Details}
We employ the Seismic Propagation through Active Regions and Convection (SPARC) $6^{\rm th}$ order linear MHD code of \citet{Han07aa}, with uniform $B_0=500$ G or 1 kG magnetic fields of inclination $\theta$ from the vertical threading the Convectively Stabilized solar Model (CSM\_B) atmosphere of \citet{SchCamGiz11aa}. 
The `observational height' for vertical velocities used in seismic calculations is 0.3 Mm, above the $a=c$ equipartition layer for both 500 G and 1 kG (0.24 Mm and 0.08 Mm respectively). A stochastic driving plane is placed at depth $z=-0.156$ Mm. The random sources produce a solar wave power spectrum with maximum power in the range 2 -- 5.5 mHz, peaking at about 3.2 mHz. Being a linear code, the overall velocity normalization is arbitrary. The standard $\fract{2}{3}$ dealiasing rule is applied to avoid spectral blocking, with $|k_x|$ and $|k_y|$ limited to about 1.9 $\rm Mm^{-1}$.

A $140\,{\rm Mm}\times140\,{\rm Mm}\times26.53\,{\rm Mm}$ box is spanned by a $128\times128\times265$ grid of spacings $\Delta x=\Delta y=1.09$ Mm covering heights $-25\,{\rm Mm} \le z \le 1.53\,{\rm Mm}$ with nonuniform $\Delta z$. Periodic lateral boundaries and absorbing PML layers at top (starting at $z_t=1.26$ Mm)
and bottom complete the definition of the computational region. 

Without loss of generality, the field lines are assumed to lie parallel to the $x$-$z$ plane. As with most such codes, SPARC normally employs an Alfv\'en speed `limiter' or `controller' (or similar) in the atmosphere to avoid infeasibly small time steps required to satisfy the CFL numerical stability condition \citep[\emph{e.g.}.][]{Han08aa,RemSchKno09aa,CamGizSch11aa}. These commonly cap the Alfv\'en speed at 20--60 $\rm km\,s^{-1}$. We do not impose such a limiter though, as it corrupts the very processes of fast wave reflection and conversion that we seek to explore \citep{MorCal13aa}. This is why we adopt lower box heights than usual, so as to keep the peak Alfv\'en speed manageable. Numerical experiment has allowed us to select box heights sufficient to encompass fast-to-Alfv\'en conversion without mandating impossibly small time steps. For the 500 G case we use $\Delta t=0.1$ s, and half that for 1 kG.

Data cubes are produced from the simulations consisting of vertical velocity values at each grid point at observation height $z_{\rm obs}=0.3$ Mm over typically $8\fract{1}{2}$ hours.  

Filtering is applied in Fourier space, both frequency and wavevector. This is more efficient than repeating the simulation for a large number of monochromatic drivers. Wavevector filtering uses a circular Gaussian ball filter with standard variation $\sigma=0.1$ $\rm Mm^{-1}$ centred at a particular horizontal wavevector $\boldk=\boldk_h$ oriented angle $\phi$ from the $x$-direction. Horizontal wavenumber $k_h=|\boldk_h|=0.5$, 0.75, or 1.0 $\rm Mm^{-1}$ determines the skip distance and phase speed, and $\phi$ represents the wave propagation direction. Frequency filtering is centred on 3 mHz and 5 mHz with a 0.5 mHz standard deviation. 

   \begin{table}
     \caption{Approximate phase speeds and skip distances associated with each of the three selected horizontal wavenumbers $k_h$ in the quiet solar model.  \label{tab:skip}}
     \begin{tabular}{ccc}
     \hline
    $k_h$ ($\rm Mm^{-1}$) & $v_{ph}$ ($\rm km\,s^{-1}$) & Skip Dist (Mm)\\
    \hline
1.0\phantom{0}   &  21.8 &  14.7 \\
0.75  &  29.3  &    19.5 \\
0.5\phantom{0}   &  43.8  &    38.8 \\
     \hline
     \end{tabular}
     \end{table}

The resultant filtered data cubes are analysed in two ways. First, the acoustic (slow) and magnetic (Alfv\'en) wave energy fluxes are calculated at $z_f=1.2$ Mm, just below the PML layer, and plotted as contoured functions of field inclination $\theta=0^\circ$, $10^\circ$, $20^\circ$,  \ldots, $90^\circ$ and wave orientation $\phi=0^\circ$, $5^\circ$, $10^\circ$, \ldots, $180^\circ$ for each frequency (3 mHz and 5 mHz). The fast wave is evanescent, and so contributes no flux. Similarly, Time-Distance 
travel time perturbations relative to quiet Sun (same atmospheric model but no magnetic field) are also plotted against $\theta$ and $\phi$.\footnote{Travel times are calculated in the standard TD manner \citep{CouBirKos06aa}, after transforming the ball-filtered data back to physical space.} These are displayed and analysed in Section \ref{results}.

\section{Results}\label{results}
In this section, wave energy fluxes at the top of the computational box will be compared with `travel time' perturbations relative to the quiet Sun for a range of magnetic field inclinations $\theta$ and orientations $\phi$, field strengths (500 G and 1 kG), horizontal wavenumbers ($k_h=1$, 0.75, and 0.5 $\rm Mm^{-1}$), and wave frequencies (3 mHz and 5 mHz). 

\subsection{Wave Energy Fluxes}\label{fluxes}

\begin{figure*}
   \centering
\includegraphics[width=.497\hsize]{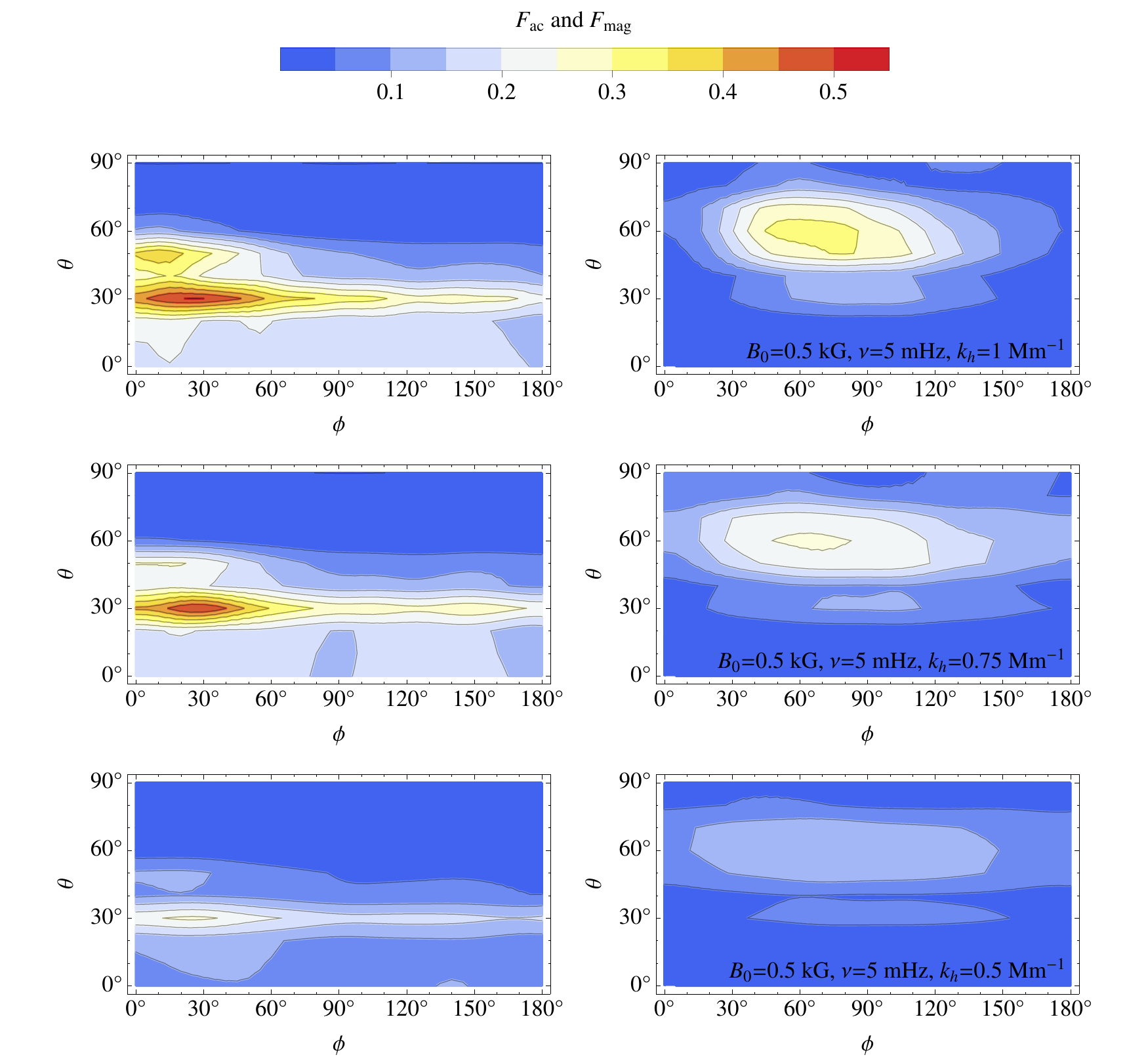}
\includegraphics[width=.497\hsize]{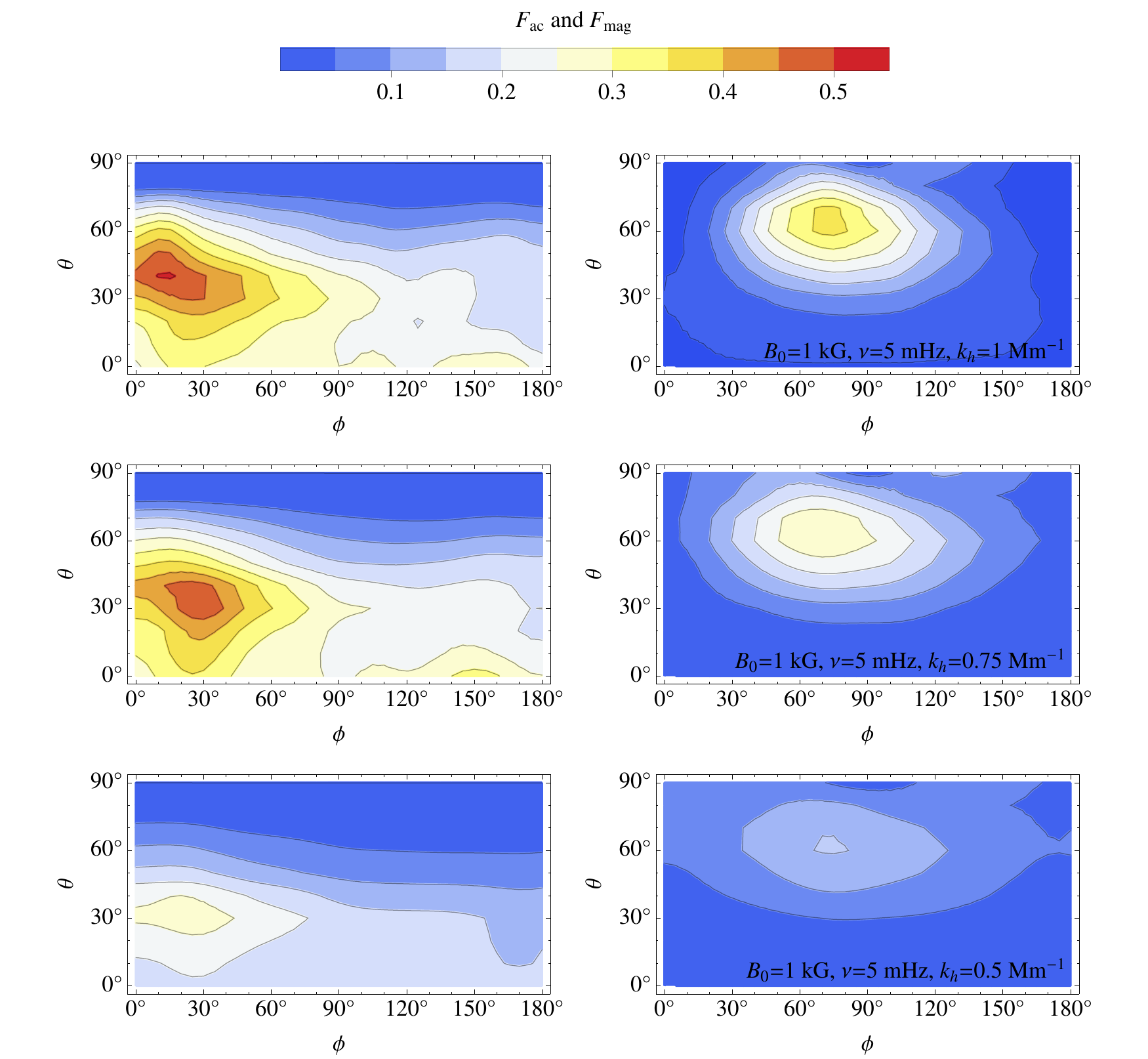}\\[6pt]
\includegraphics[width=.497\hsize]{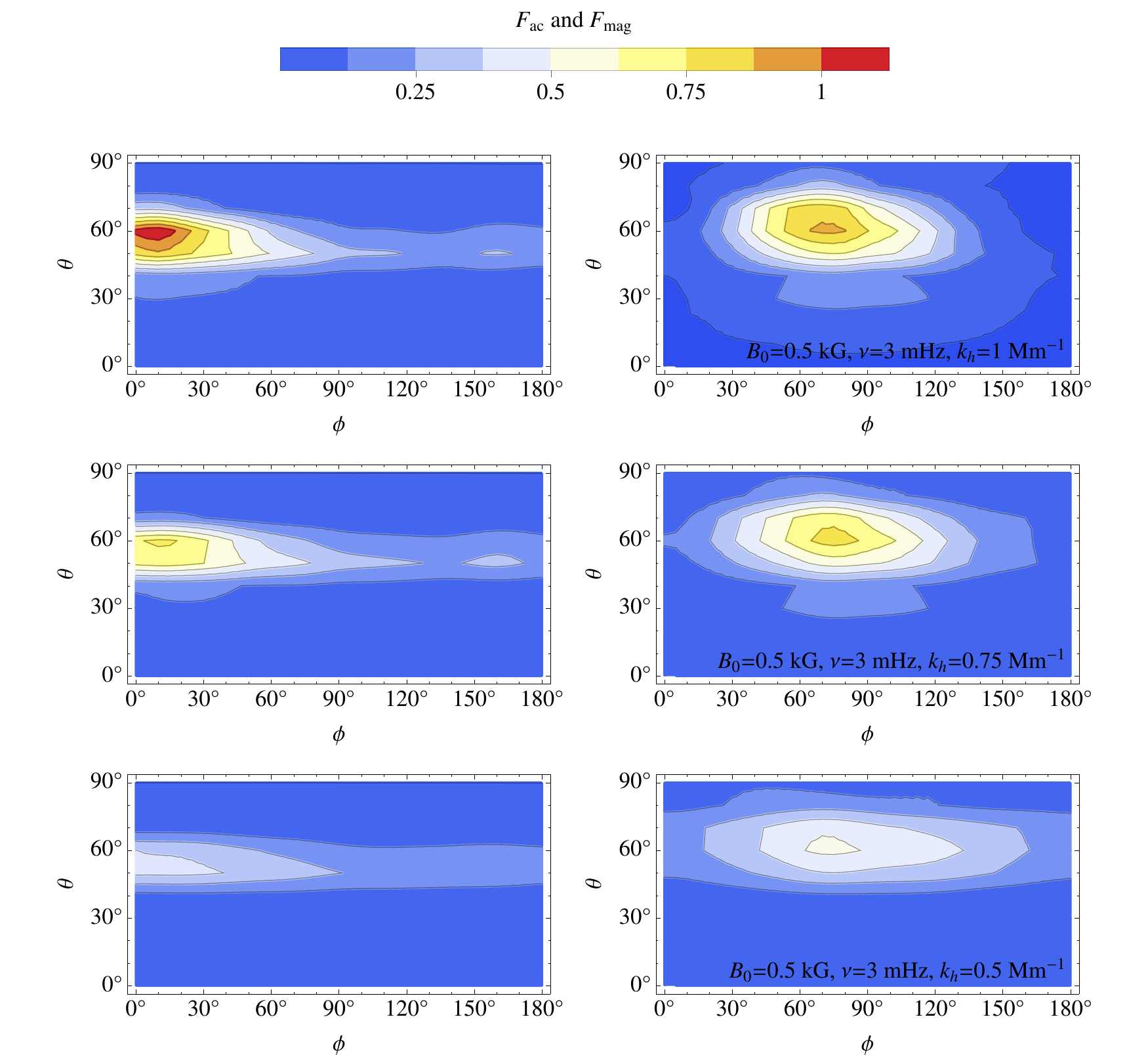}
\includegraphics[width=.497\hsize]{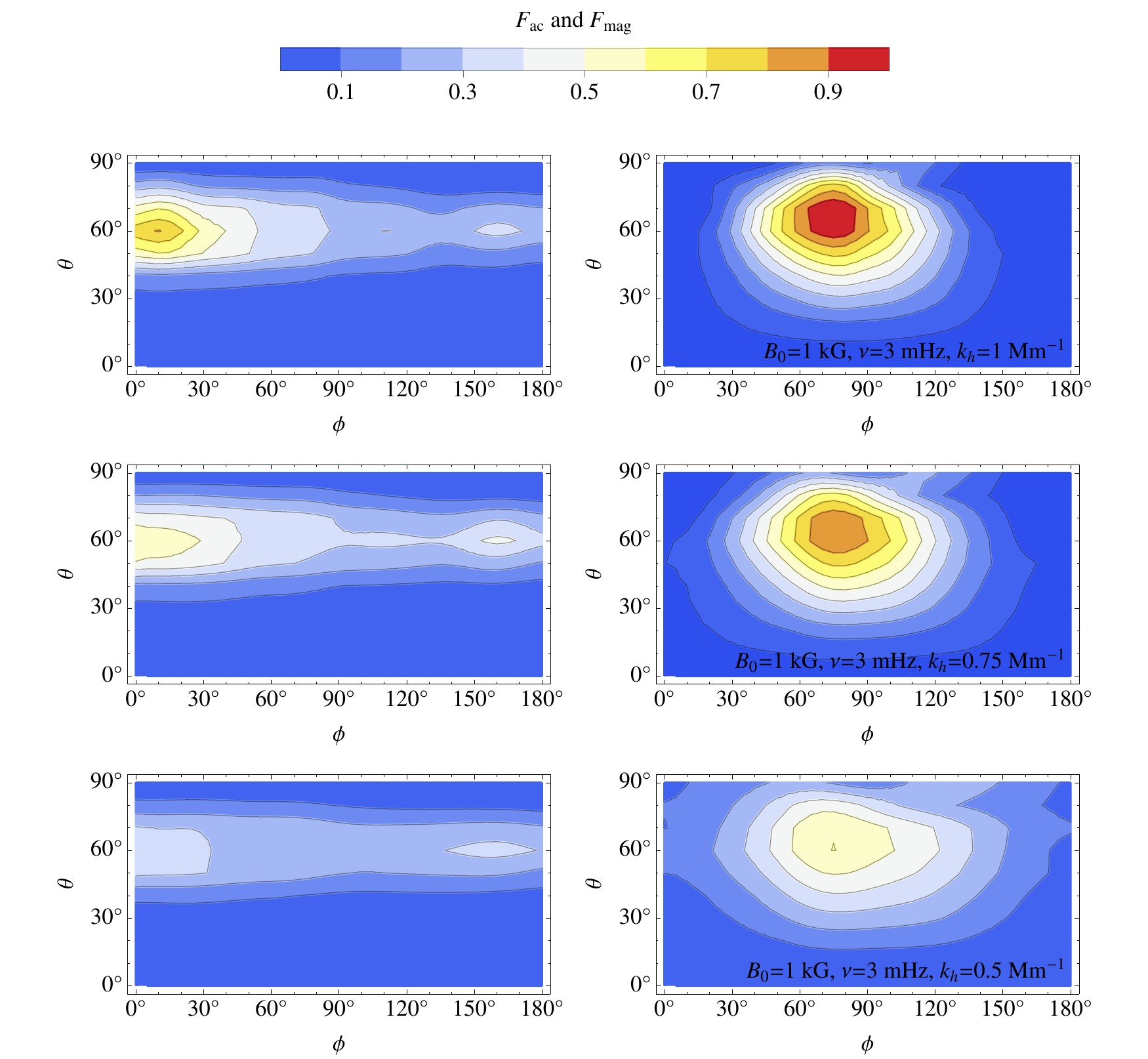}
   \caption{Four blocks of vertical wave energy flux plots against field inclination $\theta$ and orientation angle $\phi$, measured at $z_f=1.2$ Mm.
These relate respectively to the cases: $B_0=500$ G, $\nu=5$ mHz (top left); $B_0=1$ kG, $\nu=5$ mHz (top right); $B_0=500$ G, $\nu=3$ mHz (bottom left); $B_0=1$ kG, $\nu=3$ mHz (bottom right). Within each block, the first column is the acoustic flux $F_{\rm ac}$ and the second column is the magnetic flux $F_{\rm mag}$. The three rows in each block are for $k_h=1$, 0.75, and 0.5 $\rm Mm^{-1}$ respectively. 
   }
   \label{fig:Fz}
\end{figure*}

With magnetic field strengths of 500 G and 1 kG, Figure \ref{fig:Fz} depicts the top vertical acoustic and magnetic wave energy fluxes at 3 mHz and 5 mHz as functions of field inclination from the vertical $\theta$ and wave orientation as selected by the wavevector space ball filter $\phi$. The (vector) fluxes are calculated according to
\begin{equation}
\F_{\rm ac}=\langle p\,\boldv\rangle\,, \quad \F_{\rm mag}=\langle \boldb\vcross(\boldv\vcross\B_0)\rangle/\mu_0\,,  \label{flux}
\end{equation}
where $\boldv$, $p$, and $\boldb$ are the Eulerian velocity, pressure, and magnetic field perturbations, $\mu_0$ is the magnetic permeability, and the angled brackets indicate averaging over time, or over $x$ and $y$. For the most part, only vertical components of these fluxes will be plotted. Three different horizontal wavenumbers are selected: $k_h=1.0$, 0.75, and 0.5 $\rm Mm^{-1}$, corresponding to phase speeds and skip distances set out in Table \ref{tab:skip}. The overall flux normalization is arbitrary, but is consistent with the solar-like spectrum of the stochastic driving plane across the frequencies, \emph{i.e.}, having maximum power in the range 2 -- 5.5 mHz with peak at about 3.2 mHz. 

The features in Figure \ref{fig:Fz} are very much in accord with the simplified analyses of \citet{CalGoo08aa} and \citet{KhoCal11aa}. Specifically, there is negligible acoustic power at low field inclination because of the acoustic cutoff of a little over 5 mHz in the atmosphere. However, once $\omega > \omega_c\cos\theta$, acoustic waves may propagate upward (the ramp effect), and substantial flux is recorded at the top of the box. At 3 mHz, acoustic power peaks at $\theta=50^\circ$-- $60^\circ$, whilst the peak is at around $30^\circ$ at 5 mHz. These peak powers may be understood in terms of the `attack angle' $\alpha$ between the 3D wavevector and the magnetic field at the equipartition level \citep{Cal06aa,SchCal06aa,HanCal09aa}, with most efficient acoustic transmission occurring at small $\alpha$ in the generalized ray approximation. This also explains the drop-off of acoustic flux with increasing $\phi$. The slight shift in the maximum of acoustic flux away from $\phi=0$ is consistent with the 3 mHz case of \citet{KhoCal11aa}.

The magnetic flux also accords with previous (more idealized) modelling \citep{CalGoo08aa}, peaking at higher $\theta$ than the acoustic flux (especially at 5 mHz), and at wave orientations of $60^\circ$ or more. The bias toward $\phi<90^\circ$ is in accord with the schematic scenario of Figure \ref{fig:Alfschem} and the detailed cold plasma survey of \citet{CalHan11aa}. The corresponding downgoing Alfv\'en waves at $\phi>90^\circ$ are not registered by these figures. 
Strictly, there should be no Alfv\'enic flux at $\phi=0^\circ$ and $180^\circ$ because in those 2D cases the fast and Alfv\'en waves decouple. However, the finite width of the ball filter in wavevector space, the incomplete separation of the fast and Alfv\'en waves at $z_f=1.2$ Mm, and other numerical inexactitudes result in weak magnetic fluxes at those orientations. Practical time step constraints preclude us from using a taller box and higher $z_f$, where fast-to-Alfv\'en conversion is more complete, or applying magnetic fields greater than 1 kG, without invoking some form of Alfv\'en limiter.

Another feature evident in Figure \ref{fig:Fz} is increasing acoustic and magnetic flux with increasing $k_h$ in each case. This is due to the velocity power distribution in the stochastic driver, which is uniform in $\boldk_h$. That leads to RMS power in $p$ and $\boldb$ that increases linearly with wavenumber, and hence to a similar trend in flux.

\begin{figure}
   \centering
   \includegraphics[width=.75\hsize]{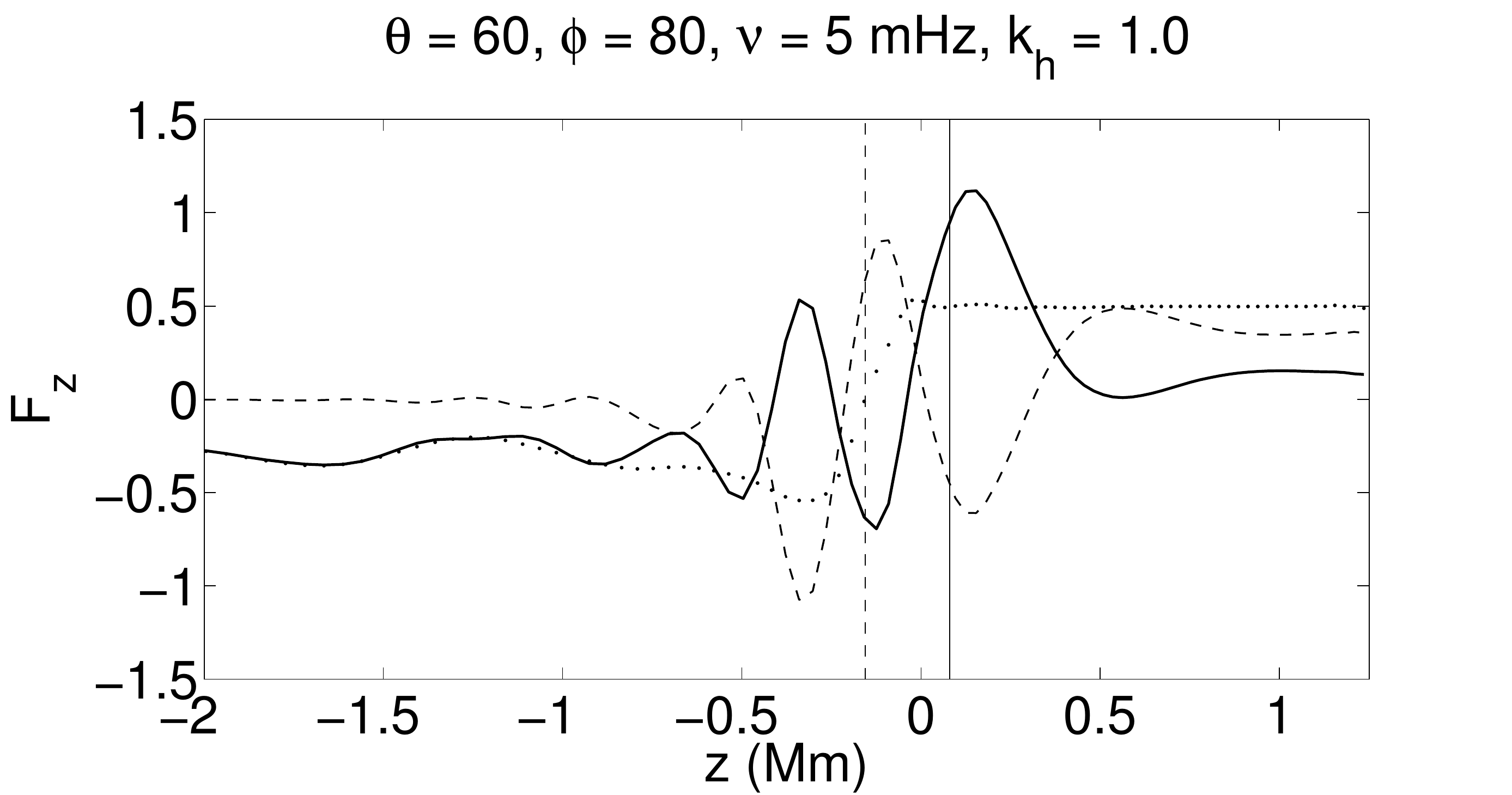}\\[6pt]
       \includegraphics[width=.75\hsize]{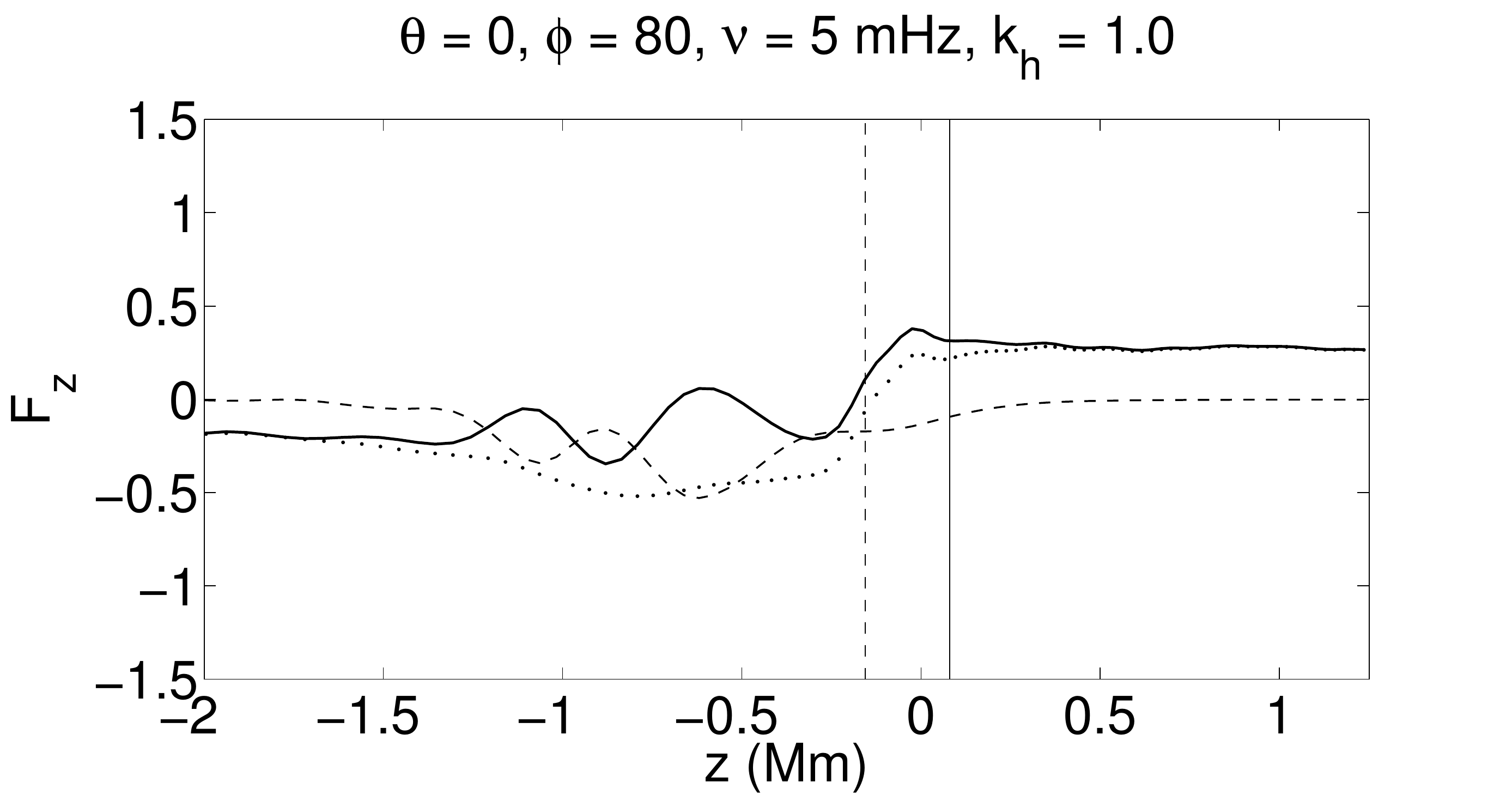} 
   \caption{Representative examples of vertical wave energy fluxes (full curve: acoustic; dashed curve: magnetic; dotted curve: total) as a function of height. The top frame pertains to $\theta=60^\circ$, $\phi=80^\circ$, $B_0=1$ kG, and $k_h=1$ $\rm Mm^{-1}$ at 5 mHz, for which significant fast-to-Alfv\'en conversion is expected (see  Figure \ref{fig:Fz}). The bottom frame is for the corresponding vertical field case $\theta=0^\circ$ (for which $\phi$ is irrelevant). The full vertical lines indicates the location of the $a=c$ equipartition height, and the vertical dashed line shows the position of the stochastic driving plane.}
   \label{fig:FzVz}
\end{figure}

Figure \ref{fig:FzVz} shows how the acoustic, magnetic, and total fluxes vary with height in two representative cases, the first with highly inclined magnetic field and the second with vertical field. Most fluxes have settled quite well to their asymptotic states by about $z_f=1.2$ Mm, indicating that our computational box is (for the most part) tall enough. There is also clearly a transfer of energy flux from acoustic to magnetic in the inclined field case over several hundred kilometres above $a=c$, as expected \citep{CalHan11aa}. This is absent in the vertical field case, as there the fast and Alfv\'en waves are decoupled.

Another reason for the (moderate) differences between the flux maps presented here and those of \citet{CalGoo08aa} is the different way that waves are injected from below. In \citeauthor{CalGoo08aa}, care is taken to inject only a pure fast wave at $z=-4$ Mm, where it is overwhelmingly acoustic. Any magnetic wave at the top of the computational box could therefore have only originated from mode transmission/conversion. The 2.5D simulations of \citet{KhoCal11aa,KhoCal12aa} similarly impose a deep fast wave driver (at $-5$ Mm). In the current SPARC simulations though, a shallow more realistic solar driver is placed at $z=-0.156$ Mm, and this excites both acoustic and magnetic oscillations directly (see Figure \ref{fig:FzVz}). Consequently, some portion of the magnetic wave energy fluxes at the top may have travelled directly along the fast and Alfv\'en dispersion relation loci rather than tunnelling from the fast branch. Nevertheless, the correspondence between the flux maps presented here and in \citeauthor{CalGoo08aa} is striking.

\begin{figure}
\begin{center}
\includegraphics[width=\hsize]{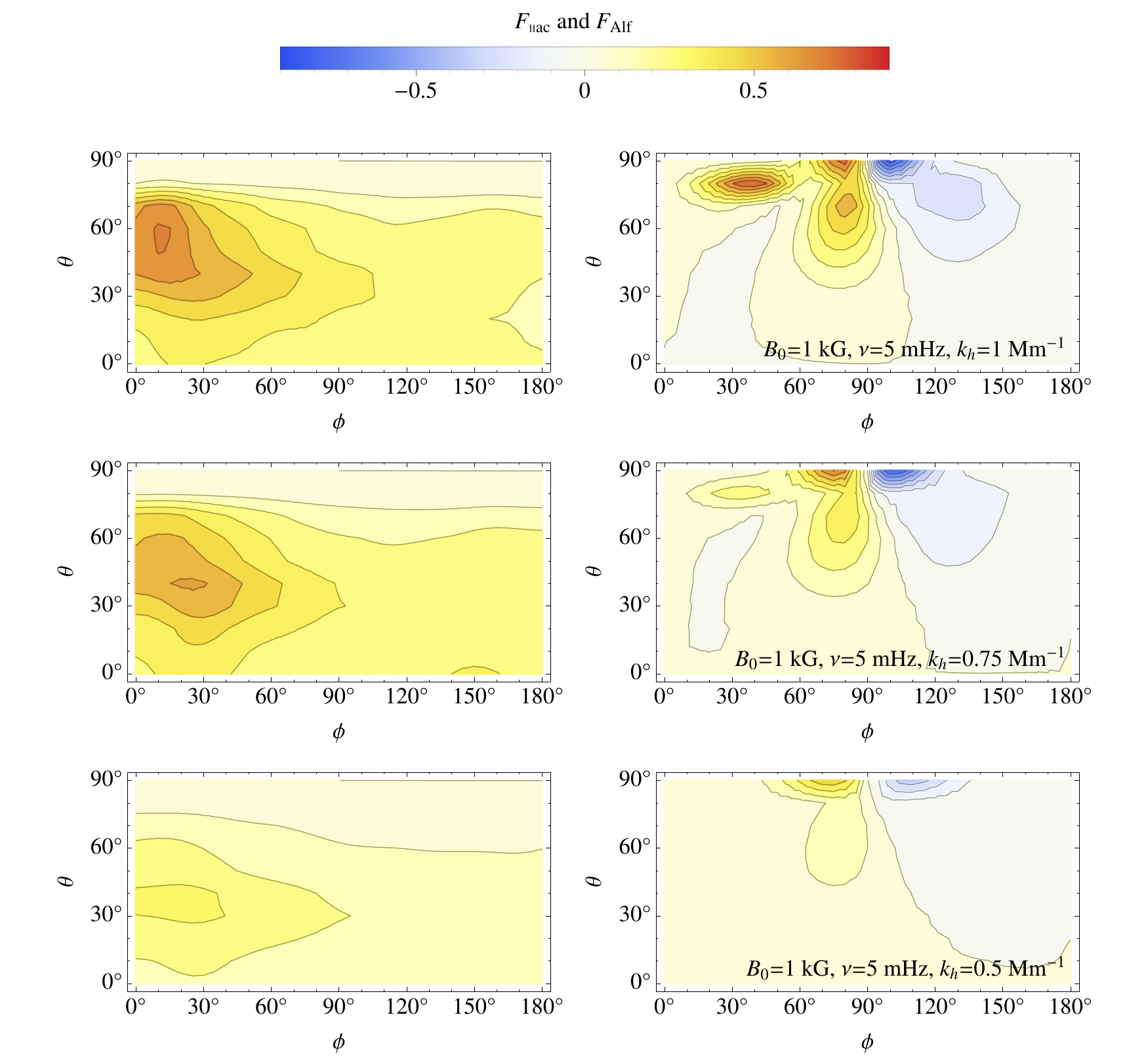}
\caption{Pure field-aligned acoustic and Alfv\'en fluxes $F_{\parallel\rm ac}$ (left column) and $F_{\rm Alf}$ (right column) for the case $B_0=1$ kG, $\nu=5$ mHz, $k_h=1$ 0.75, and 0.5 $\rm Mm^{-1}$ (top to bottom).}
\label{fig:Fpar}
\end{center}
\end{figure}

 \begin{figure*}
     \centering
     \includegraphics[width=.4\hsize]{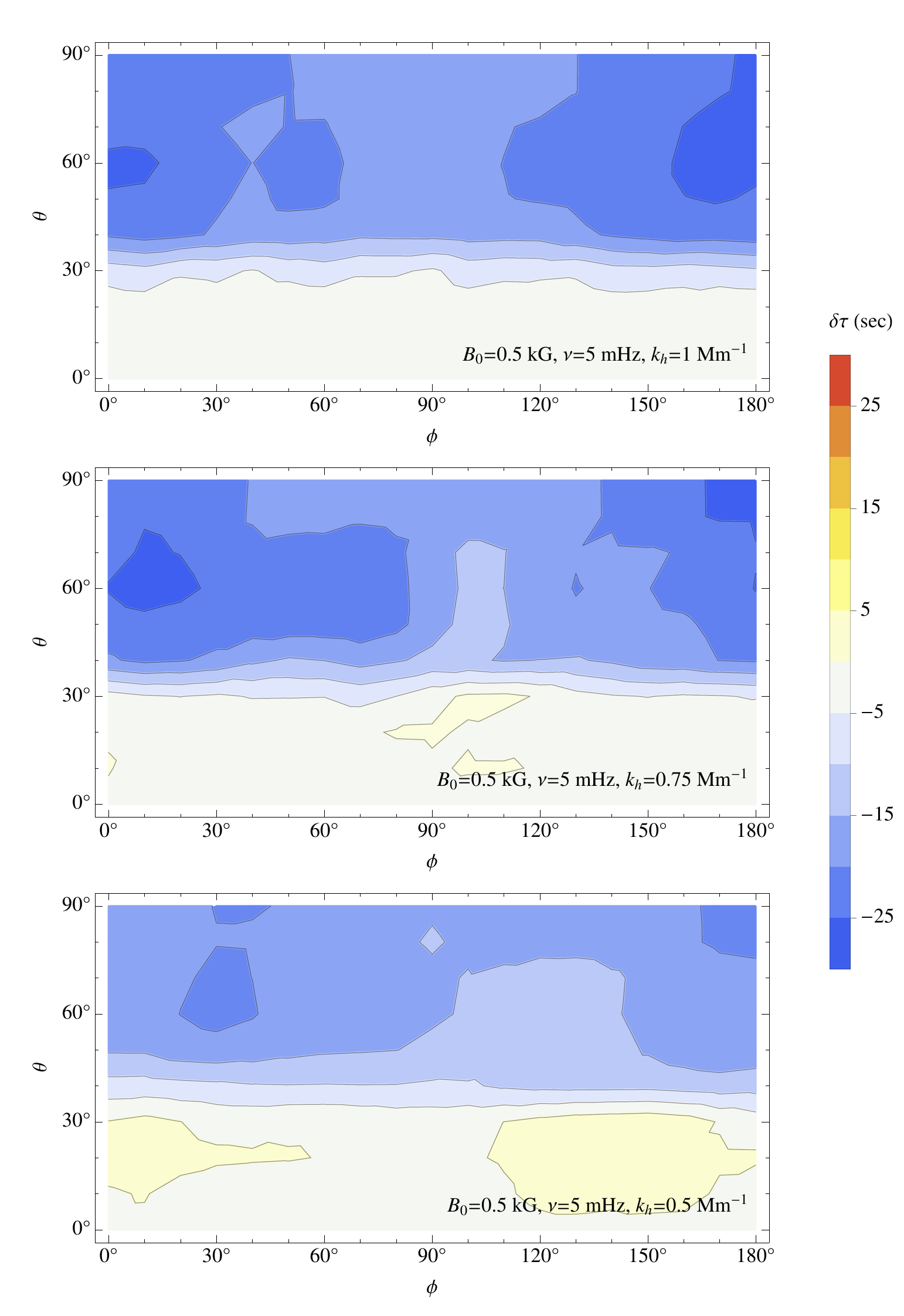} \hfil
     \includegraphics[width=.4\hsize]{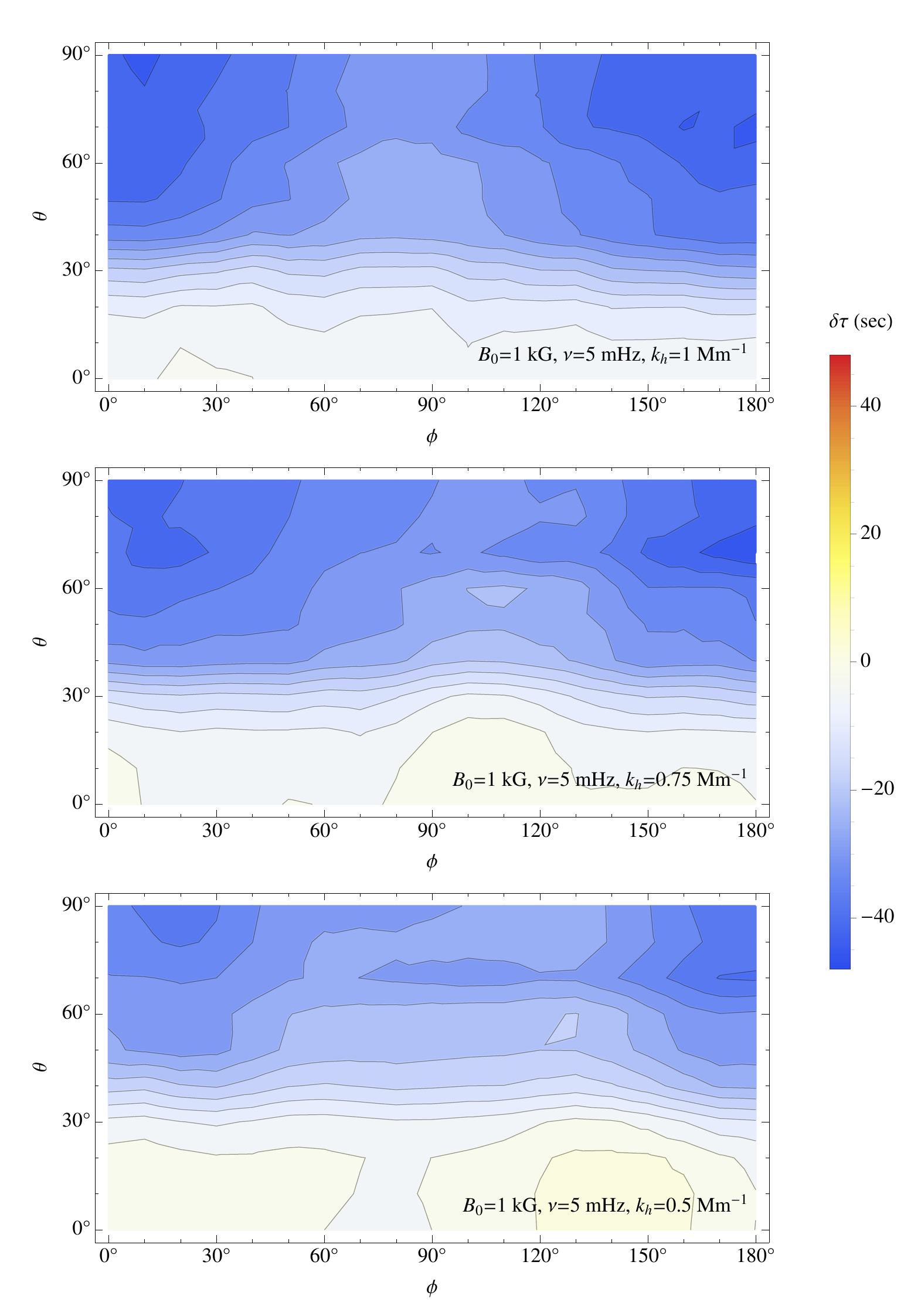}\\[8pt]
     \includegraphics[width=.4\hsize]{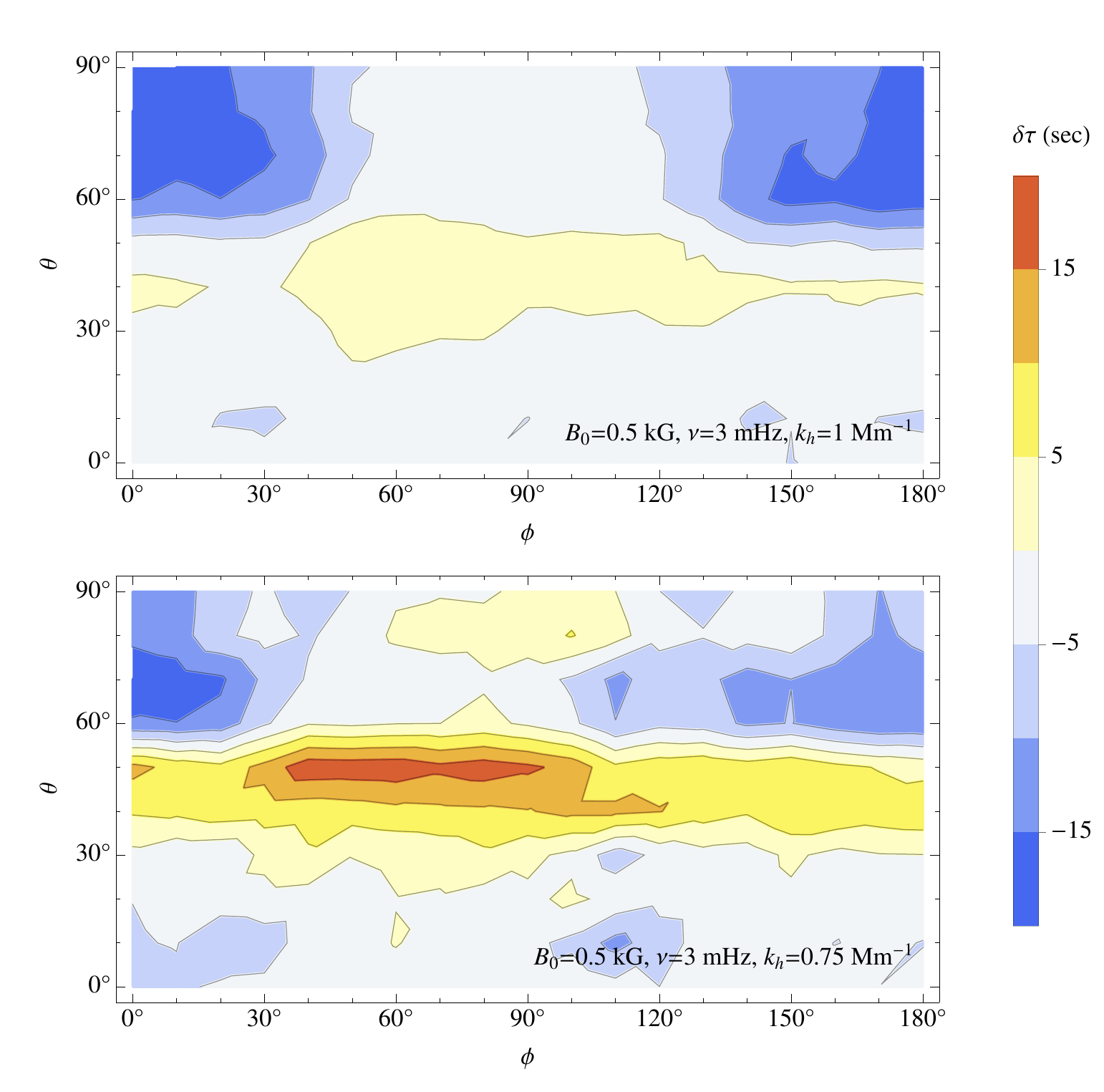} \hfil
     \includegraphics[width=.4\hsize]{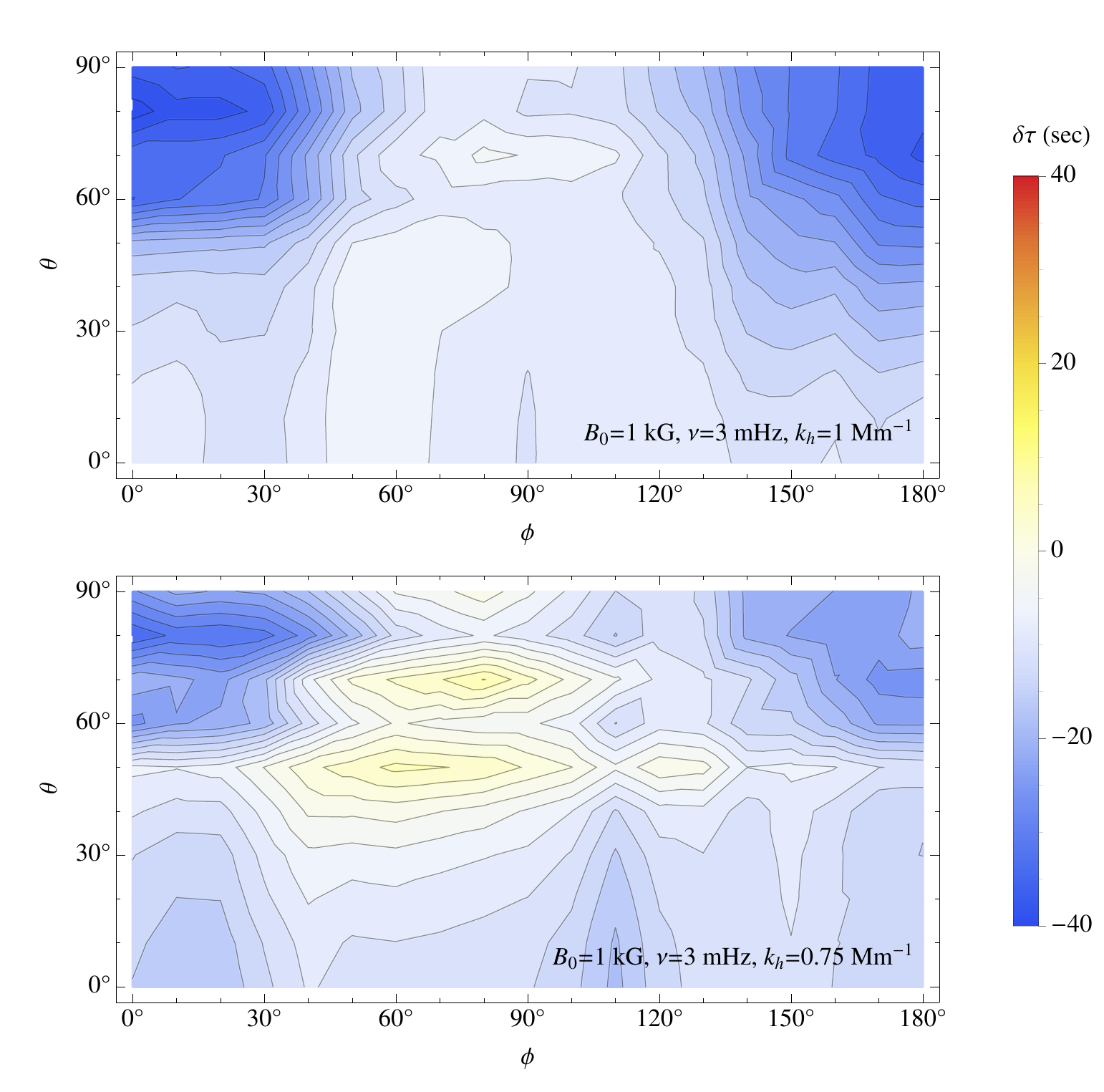}      
     \caption{Four blocks of travel time perturbations $\delta\tau$ against field inclination $\theta$ and orientation angle $\phi$. These relate respectively to the cases: $B_0=500$ G, $\nu=5$ mHz (top left); $B_0=1$ kG, $\nu=5$ mHz (top right); $B_0=500$ G, $\nu=3$ mHz (bottom left); $B_0=1$ kG, $\nu=3$ mHz (bottom right), all corresponding to the flux plots of Figure \ref{fig:Fz}. The rows in each block are for $k_h=1$, 0.75, and (for 5 mHz only) 0.5 $\rm Mm^{-1}$ respectively.}
     \label{fig:tau}
  \end{figure*}

As the magnetic field inclination $\theta$ increases toward $90^\circ$, the vertical acoustic and magnetic fluxes obviously diminish since asymptotically, as $z_f\to\infty$, both are identically field-aligned. They therefore give a biased view of energy loss rates; even for purely horizontal magnetic field, the parallel acoustic and Alfv\'en waves take energy away (horizontally) from the sites of mode conversion. Using vector identities, it is easily shown that $\F_{\rm mag} = p_{\rm mag}\boldv+\boldb\vdot\boldv\,\B_0/\mu_0$, where $p_{\rm mag}=\B_0\vdot\boldb/\mu_0$. This comprises the magnetic flux contributions to both the magneto\-acoustic and Alfv\'en waves. To select the `pure' Alfv\'en contribution, we project both $\boldb$ and $\boldv$ in the asymptotic Alfv\'en wave polarization direction 
$
\hat\E_{\rm Alf}=(\cos^2\theta\sin\phi,\cos\phi,\sin\theta\cos\theta\sin\phi)/\sqrt{1-\sin^2\theta\sin^2\phi}
$
(this is the $\e_{\rm perp}$ of \citet{KhoCal11aa} rotated into our present frame and normalized) to obtain the components $b_{\rm Alf}$ and $v_{\rm Alf}$, and identify $F_{\rm Alf}=-\langle b_{\rm Alf} v_{\rm Alf} B_0\rangle/\mu_0$. Similarly, $F_{\parallel\rm ac}=\langle p\,\boldv\vdot\hat\B_0\rangle$ is the pure field-aligned acoustic flux. Figure \ref{fig:Fpar} plots these pure parallel acoustic and Alfv\'en fluxes against $\phi$ and $\theta$. 

For horizontal field $\theta=90^\circ$ there is skew-symmetry in Alfv\'en flux about $\phi=90^\circ$, as there must be, with the Alfv\'en flux propagating in the positive direction for $\phi<90^\circ$ and the negative direction for $\phi>90^\circ$. As $\theta$ decreases though, the positive flux on $\phi<90^\circ$ is favoured since it follows the field lines sloping upward from the interaction region. On $\phi>90^\circ$ the census height $z_f$ is above the bulk of this interaction region, and so $F_{\rm Alf}$ does not yet exhibit substantial negative values -- it will do so as we move lower in the atmosphere. Indeed, if we could reliably measure $F_{\rm Alf}$ on $\phi>90^\circ$ at lower altitudes, these figures would display greater (anti-) symmetry { (see Section \ref{sec:BVP})}. The parallel acoustic flux is very similar to the vertical fluxes displayed in the top right panel of Figure \ref{fig:Fz}, with an additional $\cos\theta$ factor.


\subsection{Travel Time Perturbations}\label{TD}
`Travel time' perturbations $\delta\tau$ (relative to quiet Sun) are displayed in Figure \ref{fig:tau}, on a $10^\circ\times10^\circ$ resolution grid. They show clear manifestations of the acoustic cutoff at $\theta=30^\circ$-- $40^\circ$ for 5 mHz, and $\theta=50^\circ$-- $60^\circ$ at 3 mH where greater inclination is needed to overcome it. Below these inclinations, $\delta\tau$ is small, and typically negative, suggesting a weak speed-up due to the magnetic field. At higher inclinations, substantial travel time perturbations are seen at all orientations $\phi$, most cleanly at 5 mHz.

With only about $3\fract{1}{2}$ skips fitting into the computational box for $k_h=0.5$ $\rm Mm^{-1}$ (see Table \ref{tab:skip}), the corresponding $\delta\tau$ measurements become very noisy at 3 mHz (not shown). Even the $k_h=0.75$ results appear unreliable at this lower frequency. The large positive $\delta\tau$ at around $\theta=50^\circ$ in the $B_0=500$ G, $\nu=3$ mHz, $k_h=0.75$ $\rm Mm^{-1}$ case should therefore be regarded with some suspicion; it requires confirmation in wider boxes.

There are also clear variations in $\delta\tau$ with $\phi$ at inclinations $\theta$ sufficient for the ramp effect to take hold. In each case, negative travel time perturbations are substantial at small $\sin\phi$ but much reduced at `intermediate' $\phi$, typically around $90^\circ$. Comparison with the flux figures suggests a strong link between both acoustic and magnetic wave energy losses and travel time lags. { For magnetic field inclinations $\theta$ sufficient that $\omega>\omega_c\cos\theta$, the atmosphere is opened up to wave penetration and therefore to both types of mode conversion. With $\phi=0^\circ$, the fast magnetically-dominated waves emerging from $a=c$ return to the surface after reflection near $\omega/k_h=a$ with a different (advanced) phase compared to that of the simply reflecting acoustic waves in quiet Sun. However, if $\phi\ne0^\circ$ (or $180^\circ$), the fast wave loses more energy near its apex to the Alfv\'en wave, and seeming suffers a further phase retardation that we might interpret as partially cancelling the underlying negative travel time perturbation.

Using acoustic holography to probe sunspot penumbral oscillations, \citet{SchBraCal05aa} also detected directionally dependent phase perturbations, finding that phase shift varies with the angle between the line-of-sight and the magnetic field, with equivalent phase travel time variations with viewing angle of order 30 s. These results seem similar in type and magnitude to those found here, though it is difficult to compare precisely since line-of-sight selects fast and slow waves differently, depending on their velocity polarizations with respect to magnetic field. Their result that phase change is minimal along the line of sight is consistent with our understanding that the atmospheric slow (acoustic) wave that will be preferentially selected by that viewing angle has the same phase as the fast wave incident from below on $a=c$ \citep[there is no phase jump in the transmitted wave; see][Eqn.~(22)]{TraKauBri03aa}.
}

{
\subsection{Comparison with BVP Results} \label{sec:BVP}
\begin{figure*}
\begin{center}
\includegraphics[width=.8\hsize]{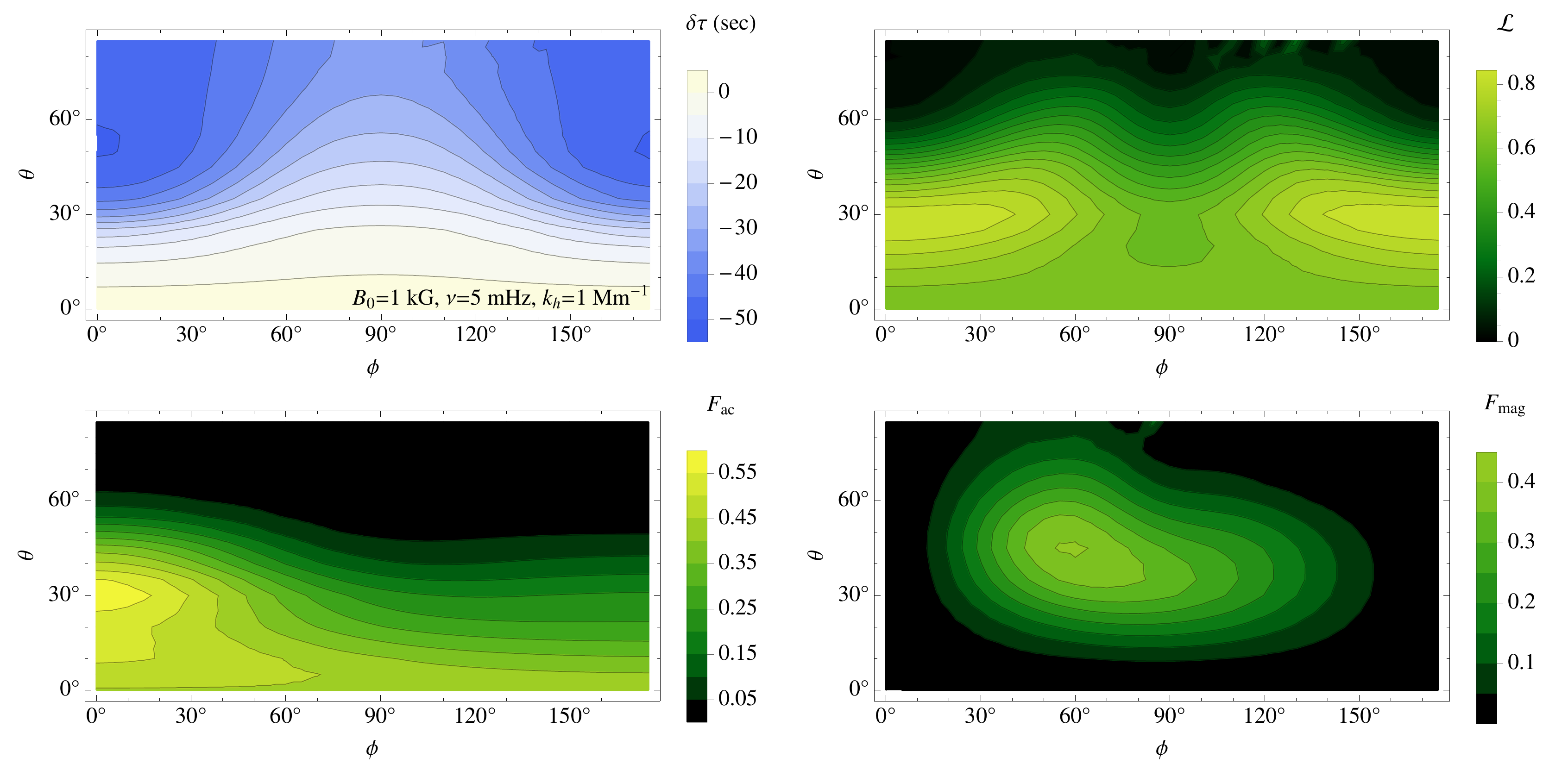}
\caption{ Results from a Boundary Value Problem calculation for $B_0=1$ kG, $\nu=5$ mHz, $k_h=1$ $\rm Mm^{-1}$. Top left: travel time perturbations; top right: total fractional vertical flux loss from the fast wave; bottom left: vertical acoustic flux, normalized by the flux of the fast wave injected at the bottom; bottom right: vertical magnetic flux, normalized by the flux of the fast wave injected at the bottom.}
\label{fig:BVP}
\end{center}
\end{figure*}

For the case of a uniform magnetic field and a horizontally invariant atmosphere, the wave propagation equations may be reduced to a $4^\mathrm{th}$ (2D) or $6^\mathrm{th}$ (3D) order system of ordinary differential equations, assuming an $\exp[\ri(k_x x+k_y y-\omega t)]$ dependence on horizontal coordinates and time. Boundary conditions may be applied so as to allow the ingress of only a pure fast wave at the bottom of the computational region (a few Mm below the solar surface), with outgoing radiation or evanescent (as appropriate) conditions on all other waves at top and bottom. This defines an ordinary differential Boundary Value Problem (BVP). Details are set out in \citet{CalGoo08aa}, \citet{Cal09ab}, and \citet{NewCal10aa}.

Figure \ref{fig:BVP} displays the results of such a calculation for one of the cases considered above: $B_0=1$ kG, $\nu=5$ mHz, $k_h=1$ $\rm Mm^{-1}$. Travel time perturbations (top left) are calculated according to the prescription of \citet{Cal09ab}. These are computed based on the phase of the reflected fast wave when it exits the box at the bottom, compared to the same for quiet Sun. No full seismic skip is constructed, but this assessment of the phase change through the surface layers amounts to the phase change per skip. It is therefore directly comparable to the time-distance results calculated above based on standard TD methodology, and indeed, the correspondence between the results of the two approaches is striking. Experiment with higher magnetic field strength (not shown) indicates that peak travel time perturbations scale roughly linearly with $B_0$, as expected from our Fig.~\ref{fig:tau} and from Fig.~7 of \citet{Cal09ab}. This is sufficient by $B_0=2$ kG for the phase difference $\delta\varphi$ to fold over the $360^\circ$ ambiguity, so that one would not be sure whether it were advanced or retarded. Since $\delta\tau=-\omega^{-1}\delta\varphi$, travel time perturbations may therefore artificially appear positive.

Figure \ref{fig:BVP} also displays the acoustic and magnetic fluxes (bottom row) calculated at the top boundary ($z=2$ Mm), though now they are normalized by the injected fast wave flux at the bottom. These are very similar to those obtained above from SPARC simulation, though the peaks of power are shifted a little in $\theta$ and $\phi$, probably because the injected flux in the simulations is not purely fast wave, since the stochastic driver is both shallow and indiscriminate in the type of waves it produces. The greater computational box height for the BVP may also contribute to the difference. It is seen that $F_{\rm ac}$ reaches over 55\% and $F_{\rm mag}$ more than 40\% in this case, showing that indeed, losses to the upper atmosphere can be substantial. 

Total fractional loss $\mathcal{L}$ is also shown (top right). This includes both upgoing and downgoing losses from the fast wave, a quantity difficult to estimate in the SPARC simulations. In other words, the remnant reflected fast wave that ultimately rejoins the interior seismic wavefield has been reduced in power by fraction $\mathcal{L}$, which can be as high as 80\%. It is notable that, as expected, $\mathcal{L}$ is more nearly symmetric about $\phi=90^\circ$ than are the upward fluxes alone. This accords with the near-symmetry of $\delta\tau$. The smaller total losses around $\phi=90^\circ$ for fixed $\theta\ga30^\circ$ are clearly reflected in $\delta\tau$. On the other hand, the small losses for $\phi\approx0^\circ$ at high $\theta$ run counter to this nexus, presumably because acoustic and magnetic losses affect phase in different ways.

In passing, we mention that the BVP travel time perturbation map for the case $B_0=0.5$ kG, $\nu=3$ mHz, $k_h=0.75$ $\rm Mm^{-1}$ (not shown) does not display the strong positive feature seen in the bottom left panel of Figure \ref{fig:tau} at mid-values of $\theta$ and $\phi$ (there is a positive peak, but it is less than 2 s). This indicates that we must be careful with very noisy TD phase fits. Longer time series or broader wave vector filters may be required.
}

\section{Discussion and Conclusions}

`Travel time' perturbations, as measured by the standard techniques of Time-Distance helioseismology, actually determine phase shifts rather than true (group) travel times. Shifts in phase trivially come about through changes in path and in propagation speed. If Fermat's principle is valid, then travel time is stationary with respect to variations of path, and so wave speed will be the main culprit. This could consist of perturbations in sound speed (temperature), magnetic field (fast wave speed), and flow (doppler shift). Flow is isolated by contrasting travel times between two points in opposite directions. There is no background flow in our simulations.

However, phase shifts are also produced by wave reflection and mode conversion \citep{Cal09aa,TraKauBri03aa}. It is a difficult (and inexact) task to theoretically determine shifts due to a single mode conversion process, but our modelled system consists of several occurring sequentially. It is further complicated by the shallow wave generation layer that excites both acoustic and magnetic waves directly, thereby differing from the model of \citet{CalGoo08aa} where care was taken to inject only acoustic waves from the bottom of the computational region { (see Section \ref{sec:BVP})}. The stochastic driving layer is however more solar-relevant, and so is of more practical relevance.

To enumerate, the mode transmissions/conversions consist of: fast-to-slow (acoustic) transmission at $a=c$; the resulting slow wave may then either escape upward or be reflected by the acoustic cutoff, depending on frequency and magnetic field inclination; the converted fast wave continues upward to be reflected near where its horizontal phase speed matches the Alfv\'en speed; conversion to upgoing and downgoing Alfv\'en waves occurs near this reflection point, depending on field inclination (with or against the wave direction); the downgoing fast wave again passes through $a=c$ and is split into fast and slow; the downgoing Alfv\'en wave impacts the surface. Each of these in turn leaves a signature on the phase. Altogether, this is far too complex to treat analytically. We are therefore left to analyse simulations by comparing wave fluxes with `travel time' perturbations. Unfortunately, the plotted fluxes do not reveal the full story. Downgoing acoustic and magnetic losses are not easily determined, and in any case are complicated by the excitation layer. Furthermore, phase shifts are not simple functions of conversion or transmission coefficients. Consequently, we should not expect a simple relationship between wave fluxes at the top of the computational domain and travel time discrepancies calculated near the photosphere. Nevertheless, there is clearly a profound link. Especially at 5 mHz, substantial $\delta\tau$ is associated with large fluxes, at field inclinations $\theta$ sufficient to allow acoustic propagation.

\begin{figure}
   \centering
   \includegraphics[width=\hsize]{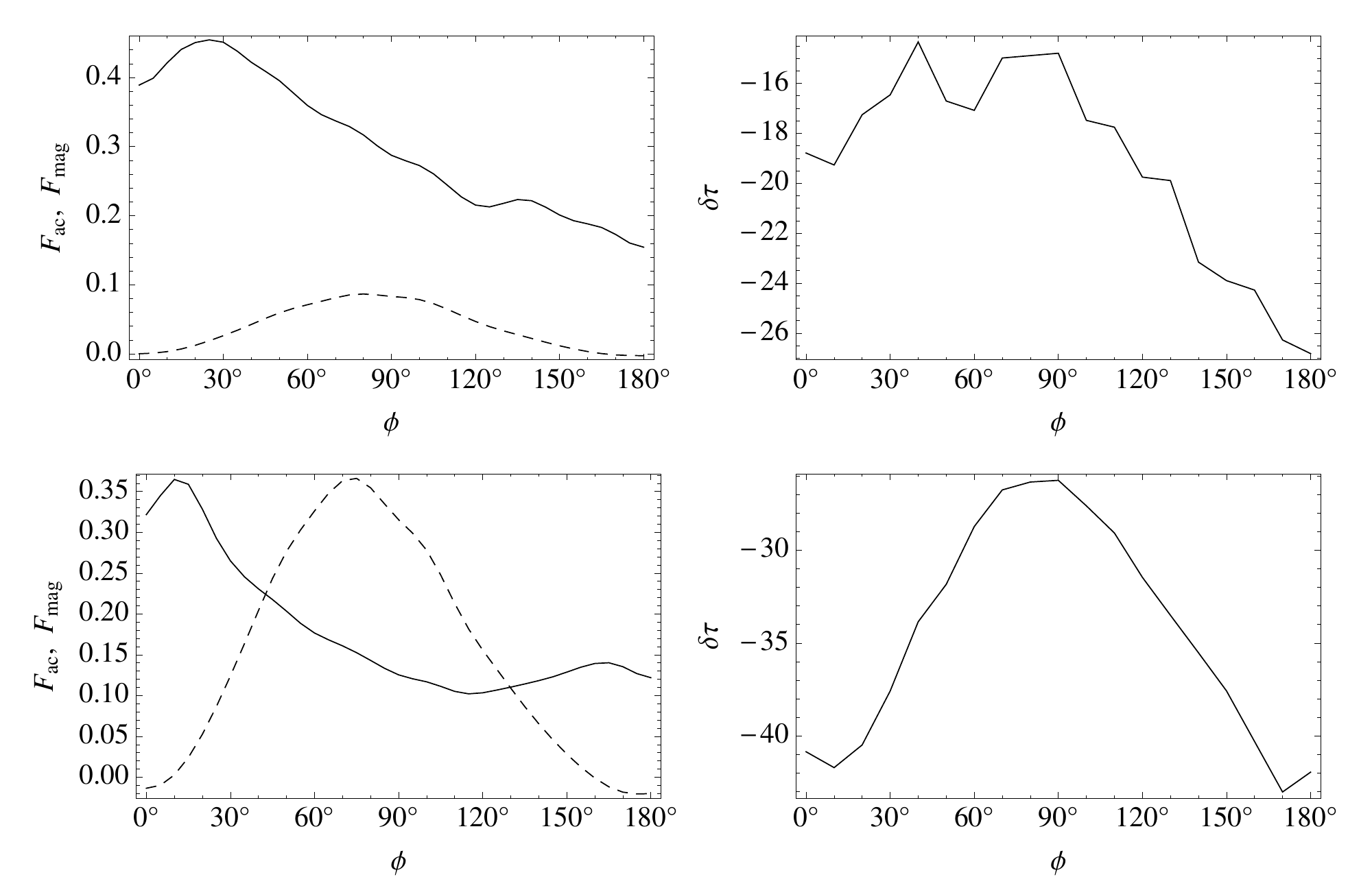} 
   \caption{Comparison of the upward acoustic and magnetic fluxes (left panels) with the travel time perturbations (right panels) against $\phi$ at $\theta=30^\circ$ (top row) and $\theta=60^\circ$ (bottom row), for the case $B_0=1$ kG, $\nu=5$ mHz, $k_h=1$ $\rm Mm^{-1}$, corresponding to the top left panels in each of Figures \ref{fig:Fz} and \ref{fig:tau}. Acoustic flux: full curve; magnetic flux: dashed curve.}
   \label{fig:compare}
\end{figure}

\begin{figure}
   \centering
   \includegraphics[width=\hsize]{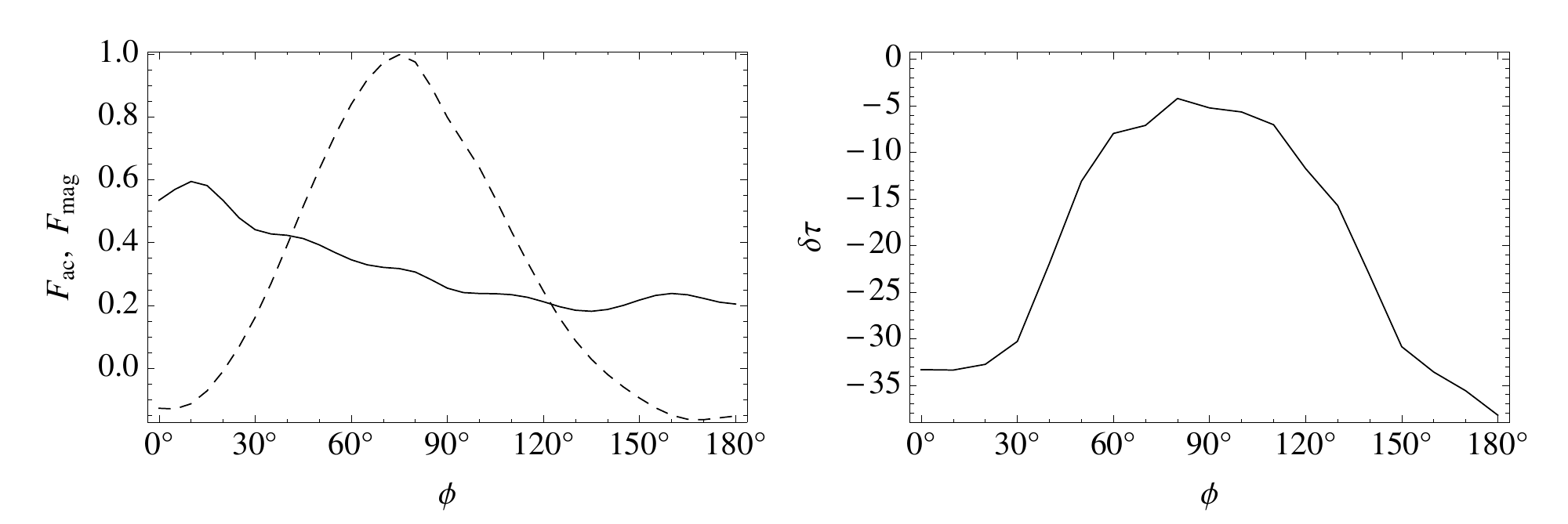} 
   \caption{Comparison of the upward acoustic and magnetic fluxes (left panel) with the travel time perturbations (right panel) against $\phi$ at $\theta=70^\circ$, for the case $B_0=1$ kG, $\nu=3$ mHz, $k_h=1$ $\rm Mm^{-1}$. Acoustic flux: full curve; magnetic flux: dashed curve.}
   \label{fig:compare3mHz}
\end{figure}

The link is perhaps clearer in Figure \ref{fig:compare}, in which the $\theta=30^\circ$ and $\theta=60^\circ$ fluxes and travel time perturbations with $B_0=1$ kG, $\nu=5$ mHz, $k_h=1$ $\rm Mm^{-1}$ are compared. For $\theta=30^\circ$, acoustic losses dominate, and produce travel time shifts of similar structure in $\phi$. Similarly, at $\theta=60^\circ$, magnetic losses dominate,\footnote{The small negative values of $F_{\rm mag}$ near $\phi=0^\circ$ and $180^\circ$ are an indication that the top of the computational box is not quite high enough for the fluxes to have attained their asymptotic values.} and now their structure is mirrored in the $\delta\tau$ plot. Similar behaviour is seen at 3 mHz in Figure \ref{fig:compare3mHz}.

{
Of course, the results of the Boundary Value Problem (BVP) calculations of Section \ref{sec:BVP} are `cleaner' than those from TD, because they are for a monochromatic wave rather than a stochastically excited spectrum of oscillations, and so do not require filtering, nor do they need to be averaged over a finite time series. The TD approach though does presage extension to more realistic atmospheres and field geometries, and importantly is directly applicable to real solar data and not just simulations. The close correspondence between the results of the two methods argues strongly for the viability of directional TD probing of real solar magnetic regions. 
}

{ Modelling of helioseismic waves is usually carried out in the linear regime, since internal oscillation velocities are invariably highly subsonic. However, our extension of the domain of helioseismology into the chromosphere poses the question of whether linearity is still a valid assumption for the extended waves addressed here. For the most part, the velocities of photospheric oscillations associated with individual p-modes are at most only a few tens of $\rm cm\,s^{-1}$ and do not of themselves grow large enough in the upper chromosphere to  be nonlinear. However, it is undoubtedly the case that, regarded as a spectrum of modes, atmospheric acoustic (slow) waves driven by the p-modes in sunspots do steepen and shock with height \citep{BogCarHan03aa}, though the relationship between the photospheric and chromospheric power spectra can be understood broadly in terms of linear theory \citep{BogJud06aa}. However, this is irrelevant to our considerations. Below the (ramp adjusted) acoustic cutoff frequency, slow waves reflect at too low an altitude for this to occur, and at higher frequencies the slow wave is lost to the seismic wavefield whether it propagates forever upward or is thermalized due to nonlinearities; in either case, the change in phase of the fast wave has already occurred at lower altitudes. By mid-chromospheric heights the plasma $\beta$ is already low and the fast wave is essentially the `compressional' Alfv\'en wave, travelling at (near) the Alfv\'en speed, which is now very large. We therefore do not expect significant nonlinearities in the fast waves. The 2D simulations of \citet{BogCarHan03aa} identify nonlinear steepening in fast waves only in the neighbourhood of a much more powerful driving piston than is relevant to our p-mode scenario.}

Overall, how are our results to be interpreted? At small magnetic field inclination, insufficient to provoke the ramp effect, both fluxes and travel time perturbations relative to quiet Sun are small, suggesting that seismic waves are largely reflected before reaching heights at which they would become involved in mode conversion. In this respect, they do not behave very differently to the quiet Sun case. However, once the ramp effect kicks in at larger $\theta$, wave paths extend into the atmosphere and both `travel times' (wave phases) and energy losses become substantial. This is so even at $\phi=0^\circ$ and $180^\circ$, where Alfve\'nic losses (essentially) vanish. Further large variations in $\delta\tau$, now positive, are then correlated with the Alfv\'enic losses as $\sin\phi$ increases. In summary, the current simulations suggest that fast-to-slow conversion at $a=c$ yields large negative travel time shifts, and that subsequent fast-to-Alfv\'en conversions produce positive shifts superimposed on and therefore partially cancelling the negative shifts. Importantly, $\delta\tau$ displays a very clear directional dependence.

A complication of the current model is that mode conversion is happening at both ends of the skip path, because the atmosphere and magnetic field is horizontally invariant. For sunspot seismology though, normally one end is in the spot and the other in a quiet Sun annular pupil where there is no conversion. This scenario will be explored in a subsequent work. For the moment though, we must realize that the travel time measurements are affected by conversions at both ends of the path, whereas the flux calculations sample only one end. This probably explains why the $\delta\tau$ contour graphs are more symmetric about $90^\circ$ in $\phi$ than are the flux plots, a feature we would not expect to persist in the sunspot/pupil model. Nevertheless, with that caveat in mind, it appears that very significant `travel time' discrepancies of several tens of seconds (depending on field strength, frequency, and wavenumber) are related to phase changes resulting from mode conversion and not true travel time changes (which is not to say that actual travel times have not changed as well due to the routing of fast and slow waves through the atmosphere). This should give pause to helioseismologists attempting to invert TD sunspot data for subsurface structure.

\vspace{2\baselineskip}\noindent
This work was supported by an award under the Merit Allocation Scheme on the NCI National Facility at the ANU, as well as by the Multi-modal Australian ScienceS Imaging and Visualisation Environment (MASSIVE). A portion of the computations was also performed on the gSTAR national facility at Swinburne University of Technology. gSTAR is funded by Swinburne and the Australian Government's Education Investment Fund.

%
\bibliographystyle{mn2e}    

\bibliography{fred}

\end{document}